\documentclass[]{spie}  

\usepackage{amsmath,amsfonts,amssymb}
\usepackage{graphicx}
\usepackage[colorlinks=true, allcolors=blue]{hyperref}

\title{Development of Solar Flare X-ray Polarimeter with Micro-Pixel CMOS Sensors}

\author[a]{Kouichi Hagino}
\author[a]{Tatsuaki Kato}
\author[a]{Toshiya Iwata}
\author[a]{Masahiro Ichihashi}
\author[a]{Hiroumi Matsuhashi}
\author[a]{Gen Fujimoto}
\author[a]{Riki Sato}
\author[b]{Hirokazu Odaka}
\author[c]{Noriyuki Narukage}
\author[a]{Shota Arai}
\author[a]{Takahiro~Minami}
\author[a]{Satoshi Takashima}
\author[a]{Aya Bamba}
\affil[a]{The University of Tokyo, 7-3-1 Hongo, Bunkyo-ku, Tokyo 113-0033, Japan}
\affil[b]{Osaka University, 1-1 Machikaneyama-cho, Toyonaka, Osaka 560-0043, Japan}
\affil[c]{National Astronomical Observatory of Japan, 2-21-1 Osawa, 181-8588, Mitaka, Tokyo, Japan}

\authorinfo{Further author information: (Send correspondence to K.H.)\\E-mail: kouichi.hagino@phys.s.u-tokyo.ac.jp}

\begin{document} 
\maketitle

\begin{abstract}
We are developing an X-ray polarimeter using micro-pixel CMOS sensors for solar flare X-ray polarimetry.
The system consists of a 2.5-$\mu$m pixel CMOS image sensor with a $12.8\times12.8{\rm ~mm^2}$ imaging area and a readout system based on a Zynq System-on-Chip.
While previous studies have validated this concept~\cite{Odaka2020}, no realistic feasibility studies have been conducted for the solar flare X-ray polarization observation.
In this work, we performed polarization sensitivity measurements at synchrotron facilities.
The results show that our polarimeter is sensitive to the X-ray polarization, exhibiting a modulation factor of 5--15\% at an energy range of 6--22~keV.
The measurements also determined the thickness of the sensitive layer to be approximately 5~$\mu$m, and the thicknesses of the insensitive layers to be 0.8 $\mu$m (Si), 2.1 $\mu$m (SiO$_2$), and 0.24 $\mu$m (Cu).
These measured thicknesses lead to a quantum efficiency of 3--4\% at 10~keV.
Based on these experimental evaluations, we estimated the sensitivity of the micro-pixel CMOS polarimeter system.
We found that, when combined with a telescope with an effective area of $\sim10 {\rm ~cm^2}$, this system can detect X-ray polarization with a polarization degree of a few percent for M-class flares.
\end{abstract}


\section{Introduction} \label{sec:intro}
X-ray polarization is revolutionizing the field of X-ray astrophysics.
In particular, the Imaging X-ray Polarimetry Explorer (IXPE), which is the first dedicated X-ray imaging-polarimetry satellite launched in 2021, has found many groundbreaking results~\cite{Weisskopf2022}.
In astrophysical X-ray sources, X-ray polarization is a probe for the magnetic field and geometrical structure.
For example, IXPE revealed the spatial structure of the magnetic field in the supernova remnant through the spatially resolved polarization measurement~\cite{Vink2022}.
The measurement of magnetic structure is very powerful, even in point sources such as the jets of supermassive black holes.
IXPE's detection of the time variability of the polarization angle of the jets indicates the shock propagation along the helical magnetic field in the jets~\cite{DiGesu2023}.
In addition, IXPE provided us with tight constraints on the geometrical structures around the black hole through its polarization degree and polarization angle~\cite{Krawczynski2022}.
Therefore, as demonstrated by IXPE, the X-ray polarization is an essential and powerful tool to understand the nature of X-ray sources.

Solar flares are the most energetic events in the solar system.
They explosively release magnetic energy, accelerating and heating plasma particles.
The resulting accelerated and heated electrons emit strong X-rays, making solar flares the brightest celestial objects in X-rays.
Compared with the other X-ray astrophysical targets, solar flares are a unique and special source where both their spatial and temporal evolution can be observed.
It is because, in general, the spatially-resolvable, diffuse X-ray sources rarely vary in observable time scale, while time-variable X-ray sources are usually point-like sources.
On the other hand, the solar flares have a spatial size of $\gtrsim 100''$ and evolve with a duration of $\sim 1000$~s, both of which can be resolved with X-ray instruments.

X-ray polarization measurements of solar flares enable the investigation of physical properties of non-thermal electrons that have not been observed before.
They can distinguish X-rays from the thermal and non-thermal electrons because only non-thermal emission is expected to be polarized.
This would enable us to determine the thermal and non-thermal energy partition in the energy release during the flare.
Additionally, the X-ray polarization is also a good probe for the transport process of accelerated particles~\cite{Jeffrey2020}.
As demonstrated in Jeffrey~et~al.~\cite{Jeffrey2020}, a higher polarization degree is expected in highly beamed electrons.
Thus, the polarization measurement provides us with the electron anisotropy, which is important to constrain electron acceleration in the solar flare.

However, solar flare X-ray polarization has not been clearly detected, though a few possible detections are reported~\cite{Suarez-Garcia2006,Zhitnik2014}.
It is partly because the solar flare X-ray polarization is expected only above 10~keV, where the non-thermal emission is dominant.
Although IXPE is one of the most sensitive instruments for X-ray polarimetry, the focal plane detector of IXPE is a gas detector, whose density is too small to detect hard X-rays above 10~keV.
Therefore, IXPE is not able to observe the solar flare X-ray polarization, and thus several future missions aimed at detecting solar flare X-ray polarization have been proposed~\cite{Savchenko2020,Fabiani2024}.

Instead of the gas detector like IXPE, we adopted the solid-state Si detector with a much higher density as a solar flare X-ray polarimeter.
In general, Si detectors have a relatively high quantum efficiency (QE), allowing the sensitive observation of high-energy X-rays.
The detection principle of our polarimeter system is the same as that used in IXPE.
In the photoelectric absorption process, the direction of photoelectrons tends to be parallel to the incident X-ray polarization due to the angular dependence of the differential cross section $d\sigma/d\Omega\propto \cos^2\phi$, where $\phi$ is the azimuthal angle of the photoelectron relative to the polarization direction.
Thus, the polarization can be measured by the electron track on the detector.
In contrast to the high detection efficiency for the hard X-rays, the detection of electron tracks is relatively difficult in Si detectors because the track length is as short as a few $\mu$m scales, requiring a very small pixel size.
Therefore, in this work, we developed a solar flare X-ray polarimeter with micro-pixel CMOS sensors with a pixel size of a few $\mu$m.

\section{Design of Micro-pixel CMOS polarimeter system}
\begin{table}[tbp]
\caption{Specifications of the micro-pixel CMOS sensor GMAX0505 and the readout board SPMU-002}
\centering
\begin{tabular}{ll | ll}
\hline\hline
GMAX0505 & & SPMU-002 &\\
\hline
Type & Front-illuminated 					& SoC & Xilinx Zynq UltraScale+ MPSoC\\
Pixel size & 2.5 $\mu$m $\times$ 2.5 $\mu$m	& CPU1 & Quad Core ARM Coretex-A53\\
Number of pixels & 25M (5120 $\times$ 5120)	& CPU2 & Dual Core ARM Coretex-R5F\\
Imaging area & 12.8 mm $\times$ 12.8 mm	& Logic cel & 256K\\
Maximum frame rate & 150 fps (10 bit), 42 fps (12 bit)& Memory & DDR4 4GB\\
\hline\hline
\end{tabular}
\label{tab:spec}
\end{table}

We developed an X-ray polarimeter system utilizing the 2.5 $\mu$m pixel CMOS sensor, GMAX0505.
Our polarimeter system is made of two components: the GMAX0505 CMOS sensor developed by Gpixel, and a Zynq-based readout board named SPMU-002 developed by the University of Tokyo and Shimafuji Electric Inc.
The basic specifications are summarized in Tab.~\ref{tab:spec}.
The sensor, GMAX0505, is a front-illuminated CMOS sensor with a pixel size of 2.5 $\mu$m, which is sufficiently small to measure the direction of the photoelectron track.
The total number of pixels is 25 Mpixels and its imaging area is $12.8\times 12.8 {\rm ~mm^2}$.

In our system, the CMOS sensor GMAX0505 is mounted on a compact sensor board with a diameter of less than 80 mm.
The electric power of $\pm 5$~V is supplied to this sensor board from SPMU-002.
The sensor is operated by the Field Programmable Gate Array (FPGA) onboard SPMU-002, and the output signal from the sensor is transferred to the FPGA in parallel through a 24-channel Low Voltage Differential Signal (LVDS) cable.
The transferred sensor data is temporally stored in the 4 GB DDR4 memory.
Then, the CPU on SPMU-002 saves the data to a data storage attached to SPMU-002 or sends it to the external computers via a Gigabit Ethernet port.
In the previous studies, utilizing the system composed of GMAX0505 and an older Zynq-based readout board named ZDAQ~\cite{Ishikawa2018}, the concept of this polarimeter system has already been established and experimentally verified~\cite{Odaka2020, Iwata2024, Tamba2024}.
However, realistic feasibility studies for the solar flare X-ray polarimetry have not yet been performed.
Thus, we evaluated the polarization sensitivity of this system based on the experimental measurements.

\section{Experiments for the Polarization Sensitivity Evaluation}
To evaluate the polarization sensitivity, the minimum detectable polarization, MDP, with 99\% significance is often used.
The ${\rm MDP}_{99}$ is ginven by
\begin{equation}
{\rm MDP}_{99}=\frac{4.29}{M\sqrt{\epsilon F_{\rm x}A_{\rm eff}T_{\rm exp}}},\label{eq:mdp}
\end{equation}
where $F_{\rm x}$, $A_{\rm eff}$, and $T_{\rm exp}$ are the X-ray source flux, effective area, and exposure time, respectively.
The other parameters, modulation factor (MF) $M$ and QE $\epsilon$, depend on the detector properties, and thus are very important to determine the sensitivity.
Here, the MF $M$ is defined as the modulation amplitude for a 100\% polarized input, and QE $\epsilon$ is the ratio of the number of detected events to those of incident X-rays.
In this section, we experimentally evaluated these parameters: $M$ and $\epsilon$.

\subsection{Modulation Factor}
\begin{figure}[tbp]
\centering
\includegraphics[width=0.45\hsize]{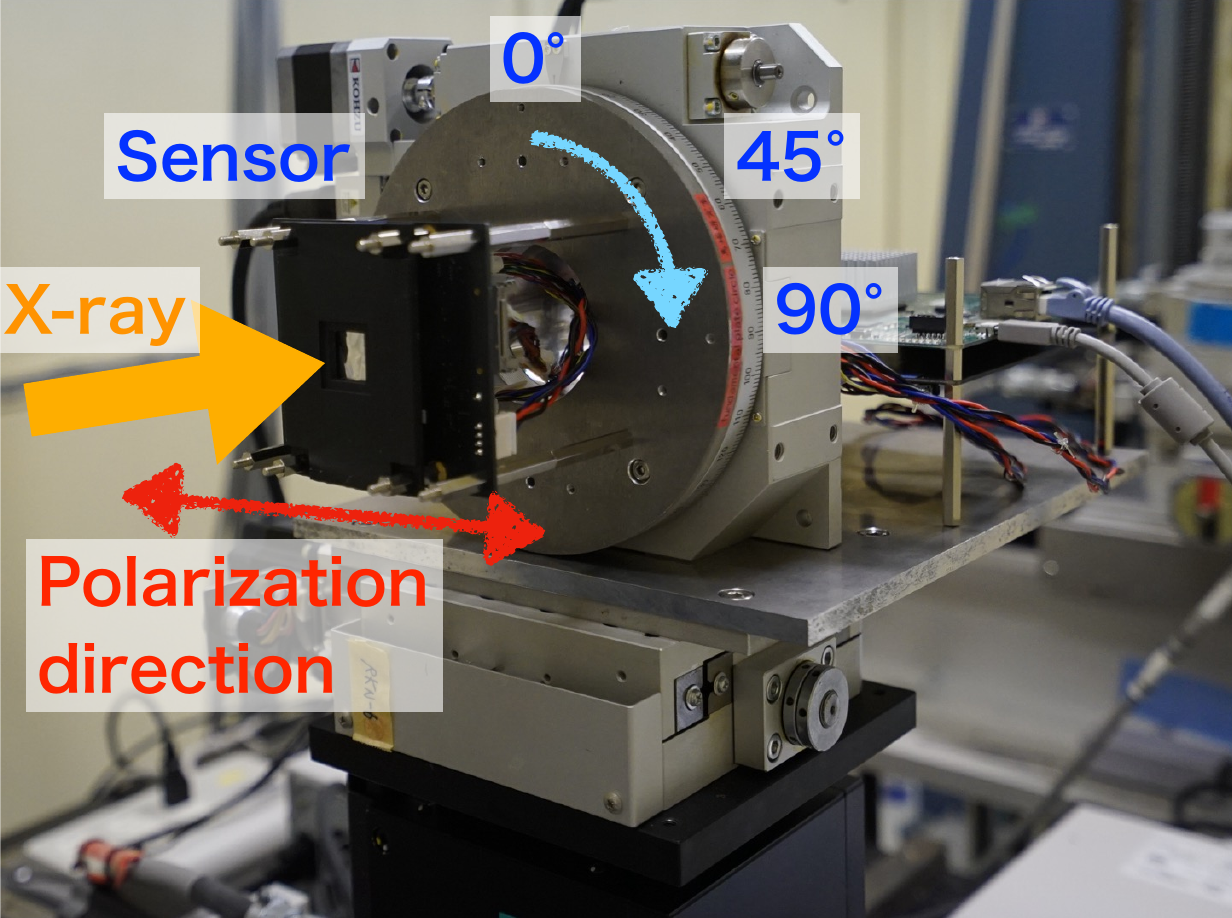}
\includegraphics[width=0.54\hsize]{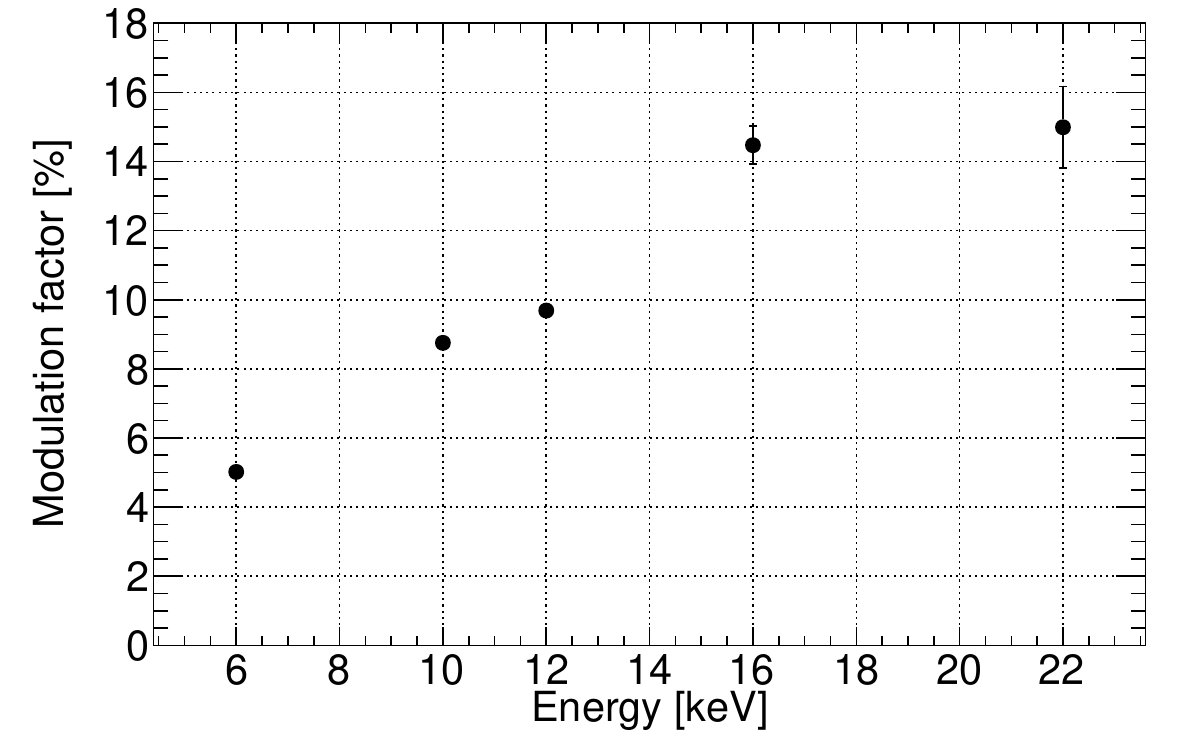}
\caption{({\it Left}) Experimental setup for the measurement of the MF at SPring-8.
({\it Right}) MF of GMAX0505 as a function of incident X-ray energy.
}
\label{fig:mf}
\end{figure}

We evaluated the MF at the medium-length beamline BL20B2 of a synchrotron radiation facility, SPring-8, in Japan~\cite{Goto2001}.
In this beamline, the incident X-rays are monochromatic and linearly polarized with a polarization degree of almost 100\% ($99.31 \pm 0.03\%$ at 12.4 keV)~\cite{Asakura2019}.
As shown in the left panel of Fig.~\ref{fig:mf}, we irradiated the X-rays, polarized to the horizontal direction, to the GMAX0505, and measured the photoelectron direction with this sensor.
The measurements were also performed for different rotation angles of $45^\circ$ and $90^\circ$ to correct the geometrically anisotropic response of the sensor.

The right panel of Fig.~\ref{fig:mf} shows the experimental results of the MF for X-ray energies ranging from 6 keV to 22 keV.
The methods used to determine the photoelectron direction and derive the MF are the same as those described by Iwata~et~al.~\cite{Iwata2024}.
The MF is smallest with $M=5\%$ for the lowest X-ray energy of 6 keV and increases up to $M=15\%$ around 20 keV, which is consistent with the previous work~\cite{Odaka2020,Iwata2024}.
Eventually, we found that the micro-pixel CMOS polarimeter is sensitive to the polarization with MFs of $M=5\textrm{--}15\%$.

\subsection{Quantum Efficiency}
The QE $\epsilon$ is essentially determined by the thicknesses of the sensitive layer and the insensitive layer.
We evaluated these thicknesses with two different methods at two different beamlines.

\subsubsection{Sensitive Layer Measurement with Reference Sensor}
\begin{figure}[tbp]
\centering
\includegraphics[width=0.49\hsize]{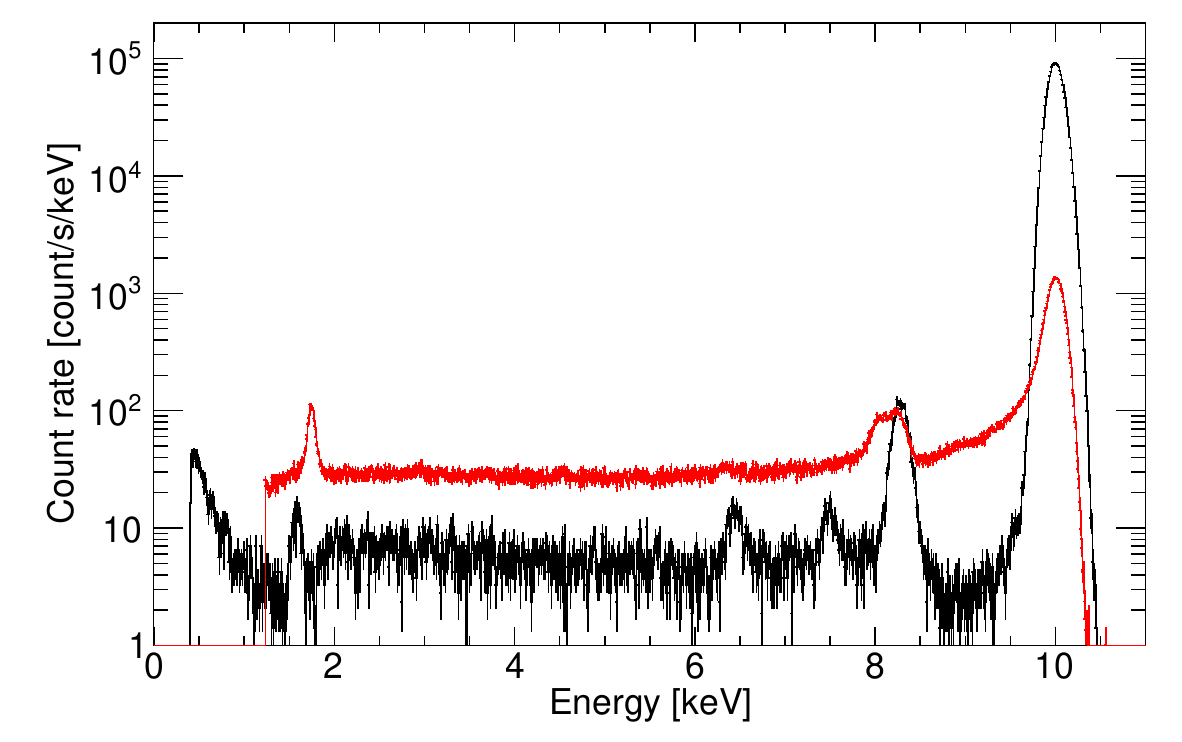}
\includegraphics[width=0.49\hsize]{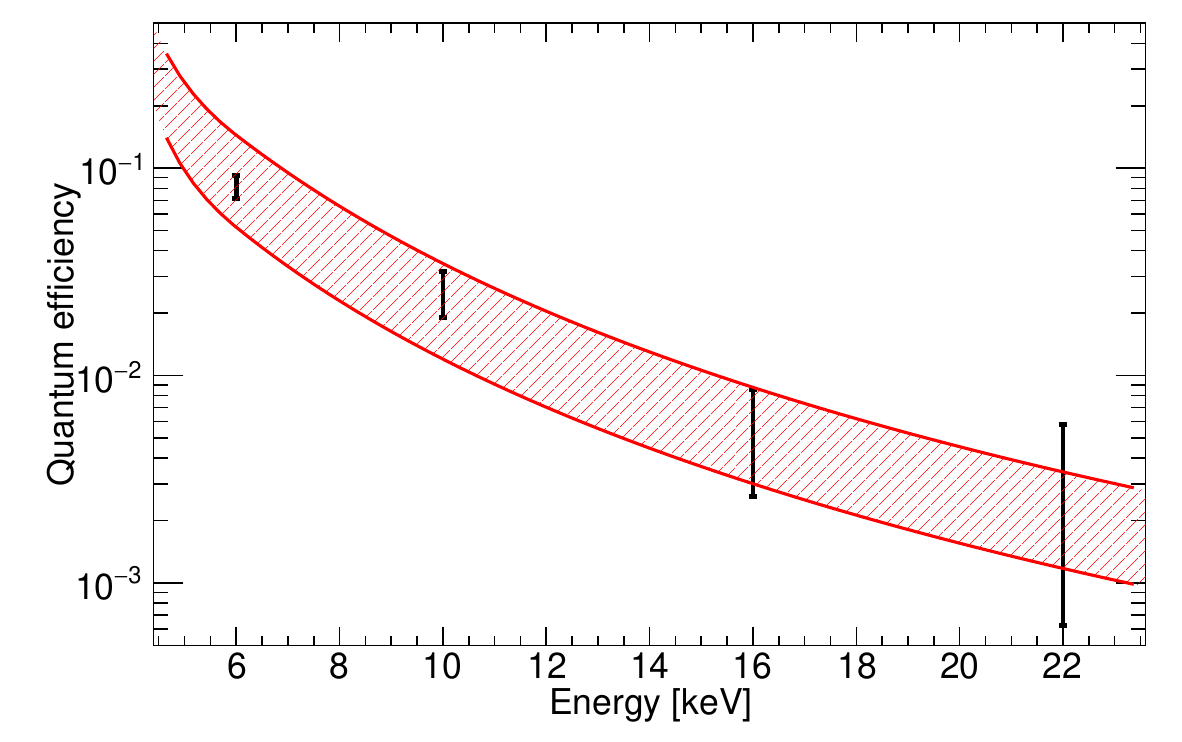}
\caption{({\it Left}) X-ray spectra at 10 keV obtained with GMAX0505 (red) and the reference SDD (black) at SPring-8.
({\it Right}) QE of GMAX0505. The red region is the best-fit QE model, considering the systematic errors.
}
\label{fig:qe_sp8}
\end{figure}

Utilizing BL20B2 in SPring-8, the same beamline as the MF measurement, we measured the thickness of the sensitive layer of GMAX0505 by comparing it with the well-calibrated reference sensor.
As the reference sensor, we used a silicon drift detector, Amptek XR-100 FAST SDD, composed of 500~$\mu$m-thick silicon with a 3.5 $\mu$m-thick B$_4$C entrance window.
In the experiment, incident X-ray flux was measured with this SDD, and the same X-ray beam was measured by replacing the SDD with GMAX0505.
The X-ray beam was collimated with a 4-mm$\phi$ pinhole made of 5~mm-thick aluminum, which is smaller than the collimator of the SDD, to avoid the uncertainty of the SDD's collimator size.

The QE was evaluated by dividing the detected count rate of GMAX0505 by that of SDD.
The major uncertainty of the measured QE in this experiment is due to the complex spectral response of GMAX0505, as shown in the left panel of Fig.~\ref{fig:qe_sp8}.
In the X-ray spectrum of GMAX0505, it is difficult to clearly evaluate how much of the low-energy structures corresponds to the signal from the incident 10 keV X-rays without detailed X-ray response studies.
Thus, we evaluated the maximum and minimum of the detected count rate by changing the energy integration range.
This corresponds to the large error bars in the right panel of Fig.~\ref{fig:qe_sp8}.
We fitted these upper and lower bounds with
\begin{equation}
f(E)=1-\exp(-\mu_{\rm Si}d_{\rm sens}),
\end{equation}
where $d_{\rm sens}$ is the thickness of the sensitive layer and $\mu_{\rm Si}$ is the X-ray absorption coefficient for Si depending on the incident X-ray energy.
As a result, we obtained the best-fit model, as shown in the red region of the figure, with a best-fit thickness of the sensitive layer ranging from 1.2 to 5.7~$\mu$m.

\subsubsection{Sensitive and Insensitive Layer Measurement with the Slant Incidence Method}
\begin{figure}[tbp]
\centering
\includegraphics[width=0.45\hsize]{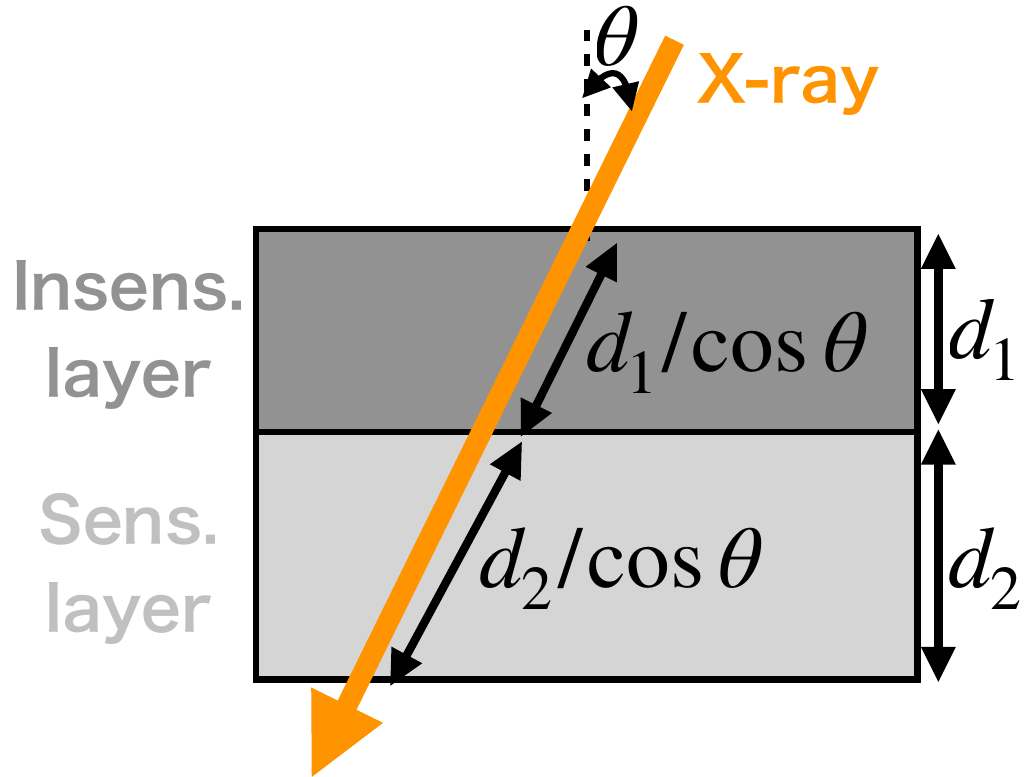}
\includegraphics[width=0.54\hsize]{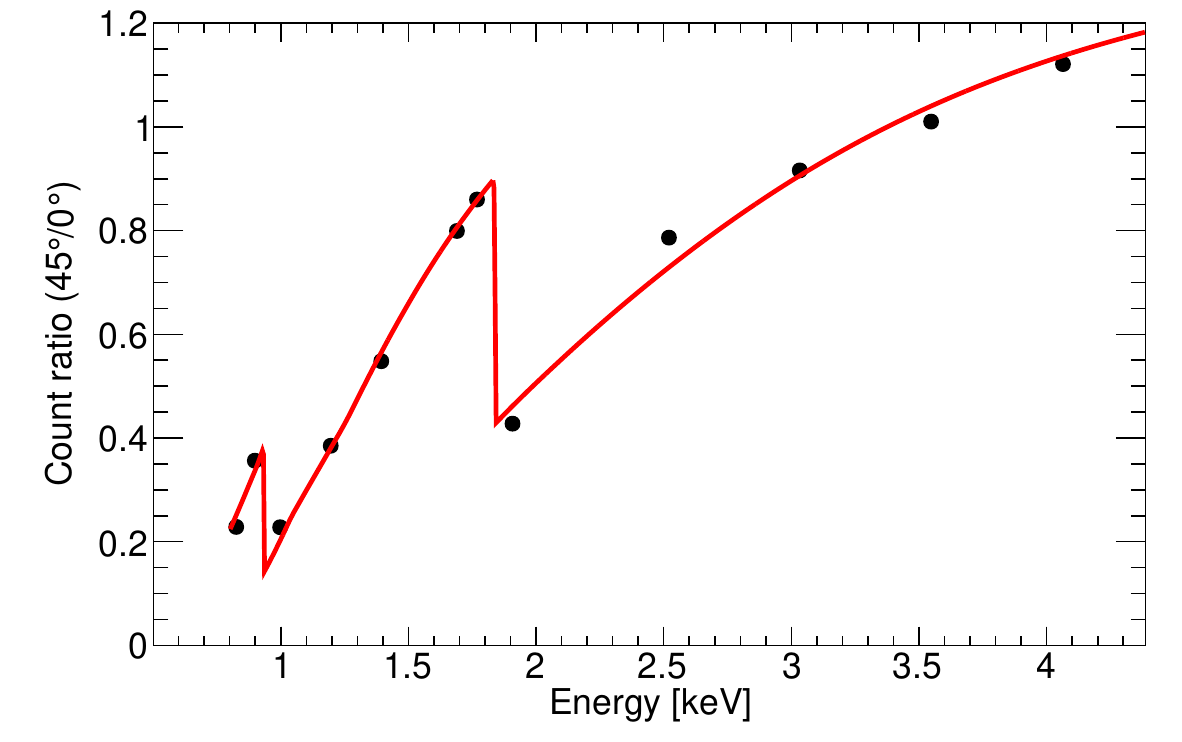}
\caption{({\it Left}) Schematic picture of the slant incidence method.
({\it Right}) Ratio of the detected count rate with an incident angle of 45$^\circ$ against that of 0$^\circ$ fitted with a 4-layer model (red; see texts for details).
}
\label{fig:qe_uvsor}
\end{figure}

In addition to the measurement using the reference sensors, we also conducted the sensitive and insensitive layer measurements with the slant incidence method~\cite{Hayashida2003}.
As shown in the left panel of Fig.~\ref{fig:qe_uvsor}, in this method, X-ray is irradiated with an incident angle $\theta$. 
If the incident angle $\theta$ is not $0^\circ$, the effective thickness of the insensitive and sensitive layers increases by a factor of $1/\cos\theta$. 
Thus, a ratio $R(E)$ of the detected count rate at an incident angle $\theta$ to the count rate at $0^\circ$ is expressed as
\begin{equation}
R(E) = \frac{\exp\left(-\sum\mu_{\rm insens}d_{\rm insens}/\cos\theta\right)}
{\exp\left(-\sum\mu_{\rm insens}d_{\rm insens}\right)}
\times
\frac{1-\exp\left(-\mu_{\rm Si}d_{\rm sens}/\cos\theta\right)}
{1-\exp\left(-\mu_{\rm Si}d_{\rm sens}\right)},
\end{equation}
and determined independently from the incident X-ray intensity.
The first factor indicates the change in insensitive layer thickness, and the second factor indicates the change in sensitive layer thickness.
The summation $\sum \mu_{\rm insens}d_{\rm insens}$ in the first factor accounts for the effect of multiple insensitive layers, where $\mu_{\rm insens}$ is the absorption coefficient and $d_{\rm insens}$ is the thickness of each insensitive layer.

We performed this experiment at another synchrotron radiation facility, UVSOR, in Japan.
We used a soft X-ray beamline BL2A in UVSOR, which serves monochromatic X-rays at energies from 0.6 keV to 4.0 keV~\cite{Hiraya1992}.
This soft X-ray band is sensitive to the thickness of the insensitive layer because of its relatively large absorption coefficient.
Since these soft X-rays are strongly affected by the atmospheric attenuation, the GMAX0505 sensor was installed in a vacuum chamber.
The sensor was connected to a goniometer attached to one of the chamber's ports, allowing the sensor to be rotated relative to the incident X-rays.
We irradiated X-rays with two incident angles of $0^\circ$ and $45^\circ$, and evaluated the count rate for each incident angle.

As a result of the experiment, we obtained the count ratio between $45^\circ$ and $0^\circ$ as shown in the right panel of Fig.~\ref{fig:qe_uvsor}.
Clearly, we can see the copper (Cu) L-edge at 0.93 keV and the silicon (Si) K-edge at 1.84 keV, indicating the presence of Cu and Si in the insensitive layer.
Since Cu is often used as a metal layer in recent CMOS circuits, it is reasonable that it is present in the CMOS readout circuits implemented on the irradiated side (front side in GMAX0505).
In addition to Cu and Si, we assumed an additional SiO$_2$ layer as the insensitive layer to reproduce the measured Si K-edge depth while keeping the overall attenuation level.
This is also reasonable because SiO$_2$ is a common insulator material used in CMOS circuits.
By combining these, we adopted a 4-layer model composed of a sensitive Si layer and 3 insensitive layers with Si, SiO$_2$, and Cu.
With this model, the experimental data were successfully reproduced as shown in the right panel of Fig.~\ref{fig:qe_uvsor}.
Considering the systematic uncertainty in the X-ray incident angle of $\pm1^\circ$, we obtained the best-fit thicknesses of $6.2\pm2.0{\rm ~\mu m}$ (sens. Si), $0.8\pm0.4{\rm ~\mu m}$ (insens.~Si), $2.1\pm0.4{\rm ~\mu m}$ (SiO$_2$), and $0.24\pm0.04{\rm ~\mu m}$ (Cu).

\subsubsection{Quantum Efficiency}
\begin{figure}[tbp]
\centering
\includegraphics[width=0.7\hsize]{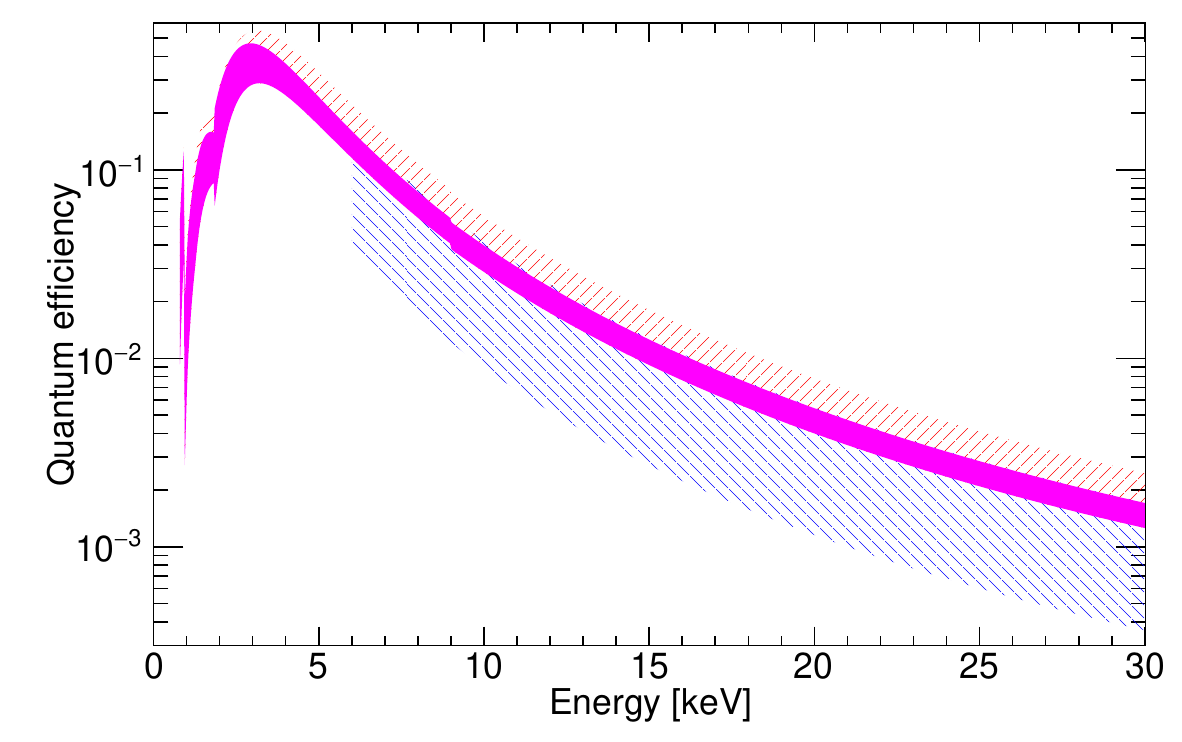}
\caption{Estimated QE based on two experiments at SPring-8 and UVSOR.
The orange and blue shaded regions are QEs based on experiments at SPring-8 and UVSOR, respectively.
The magenta region indicates the best estimation of QE, where two experimental results overlap.
}
\label{fig:qe}
\end{figure}

As shown in Fig.~\ref{fig:qe}, combining the experimental results at SPring-8 and UVSOR yields the best estimate of the QE of GMAX0505.
The two sets of results overlap within their systematic uncertainties.
The interval of thickness overlap of the sensitive layer ranges from 4.2~$\mu$m to 5.7~$\mu$m, which is the most plausible value based on our experiments.
The estimated QE peaks at 3~keV with a maximum value of 30--40\% and decreases at higher energies down to 0.1--0.2\% at 30~keV.
The estimated quantum efficiency is 3--4\% at around 10~keV, which is the most important energy for X-ray polarization measurements of solar flares.

\section{Sensitivity to the Solar Flare X-ray Polarization}
We finally estimate the polarization sensitivity for the solar flare X-ray, ${\rm MDP}_{99}$, with GMAX0505 by combining the MF and QE of GMAX0505, which were experimentally evaluated at two synchrotron radiation facilities.
In Eq.~\ref{eq:mdp}, the effective area $A_{\rm eff}$, X-ray flux $F_{\rm x}$, and exposure time $T_{\rm exp}$ should be determined, except for the measured MF $M$ and QE $\epsilon$.
For estimation purposes, we assumed the use of a compact mirror similar to the one used for the FOXSI rocket~\cite{Christe2016}.
This mirror has an effective area of $A_{\rm eff}\simeq 20{\rm ~cm^2}$ at 10 keV, decreasing to $A_{\rm eff}\simeq 5{\rm ~cm^2}$ at 15~keV.
We estimated the X-ray flux $F_{\rm x}$ assuming the spectral shape presented by Jeffrey~et~al.~\cite{Jeffrey2020}, scaling it to the fluxes of the C1-, M1-, and X1-classes.
The exposure time $T_{\rm exp}$ was assumed to be $T_{\rm exp}=1000{\rm ~s}$.

\begin{figure}[tbp]
\centering
\includegraphics[width=0.7\hsize]{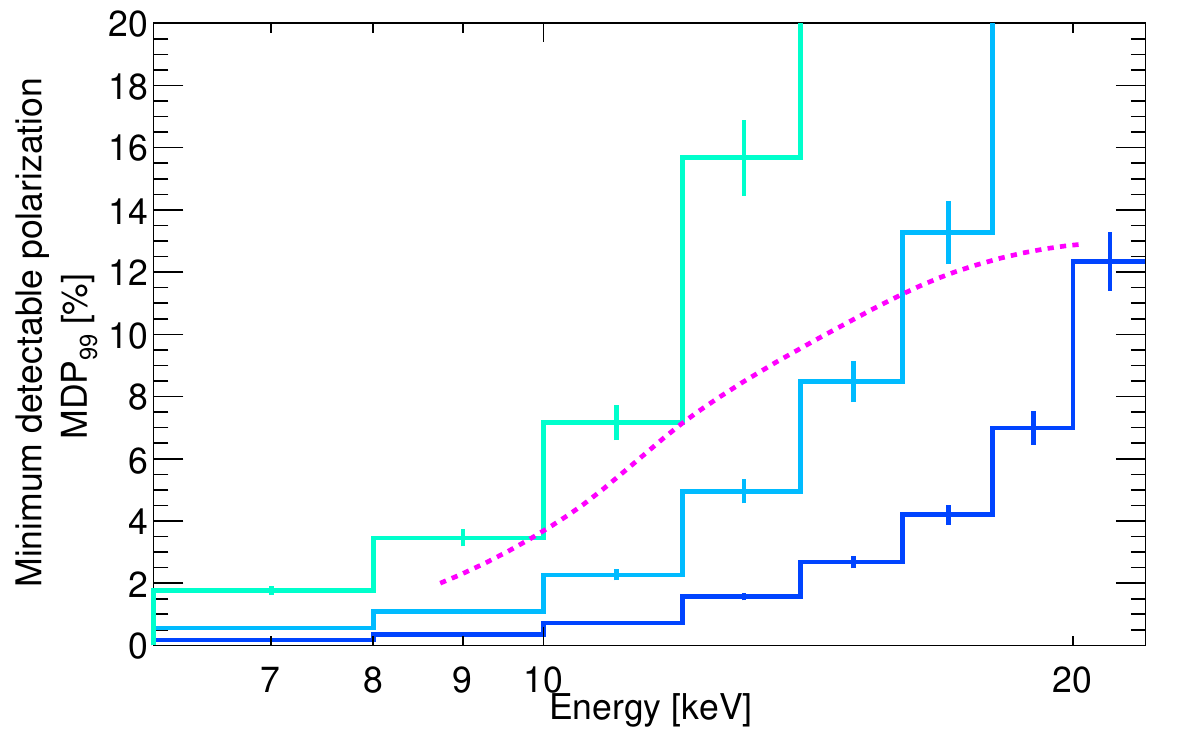}
\caption{Minimum detectable polarization ${\rm MDP}_{99}$ estimated based on our experimental measurements of MF and QE.
The light green, light blue, and blue lines are ${\rm MDP}_{99}$ for C1-, M1-, X1-class flares, respectively. 
The magenta dashed curve is the theoretically expected polarization degree based on Jeffrey~et~al.~\cite{Jeffrey2020}.
}
\label{fig:mdp}
\end{figure}

The estimated polarization sensitivity ${\rm MDP}_{99}$ is shown in Fig.~\ref{fig:mdp}.
It was calculated with an energy bin width of 2 keV.
At the 10--12 keV bin, ${\rm MDP}_{99}$ is below 1\% for X1-class flares and around 2\% for M1-class flares.
For C1-class flares, it is much worse, reaching 7\%, due to the dependence of ${\rm MDP}_{99}$ on the X-ray flux $F_{\rm x}$.
At higher energies, ${\rm MDP}_{99}$ gets worse rapidly, primarily due to the sharp decline in the incident X-ray flux, as well as lesser contributions from decreases in the QE $\epsilon$ and the effective area $A_{\rm eff}$.
For comparison, the theoretically expected polarization level from Jeffrey~et~al.~\cite{Jeffrey2020} is also overplotted in Fig.~\ref{fig:mdp}.
The estimated ${\rm MDP}_{99}$ for M1-class flares is lower at energies around 10--15~keV.
This indicates that the micro-pixel CMOS polarimeter can detect the polarization with a few \% level of polarization degree for M-class flares.

\section{Summary}
We are developing a solar flare X-ray polarimeter with micro-pixel CMOS sensors.
This polarimeter system is composed of a front-illuminated CMOS sensor GMAX0505 with a small pixel size of 2.5 $\mu$m and a Zynq-based readout board SPMU-002.
We performed experimental evaluations of the modulation factor (MF) and quantum efficiency (QE) of this system at two synchrotron radiation facilities, SPring-8 and UVSOR.
As a result, the MF was found to be 5--15\% at 6--22 keV energy range by irradiating almost fully polarized X-ray beam at SPring-8.
The QE was evaluated by measuring the thickness of sensitive and insensitive layers of GMAX0505.
By combining the results of two different methods, GMAX0505 was found to have a sensitive layer thickness of about 5~$\mu$m and 3 insensitive layers with thicknesses of 0.8~$\mu$m (Si), 2.1~$\mu$m (SiO2), and 0.24~$\mu$m (Cu).
These results lead to the QE of 3--4\% at 10~keV.
Using these measured values and assuming the use of the FOXSI mirror, we found that a few \% level of X-ray polarization is detectable for the M-class solar flare with our micro-pixel CMOS polarimeter system.


\acknowledgments 
This work was supported by JSPS KAKENHI Grant Number JP24K00638.
The synchrotron radiation experiments were performed at the BL20B2 of SPring-8 with the approval of the Japan Synchrotron Radiation Research Institute (JASRI) (Proposal No. 2024A1480, and 2023A1476).
A part of this work was conducted at the BL2A of UVSOR Synchrotron Facility, Institute for Molecular Science (IMS program 24IMS6004).
 

\bibliography{report} 
\bibliographystyle{spiebib} 

\end{document}